%% file: itadata2024.tex
\documentclass[final, runningheads]{llncs}
\usepackage[T1]{fontenc}
\usepackage{xcolor}
\usepackage{graphicx}
\usepackage{multirow}
\usepackage{rotating}

\usepackage{tikz}
\usepackage{pgfplots}
\usepackage{paralist}

\usetikzlibrary{shapes,shapes.misc,decorations.pathreplacing,arrows.meta}
\usetikzlibrary{calc}
\usetikzlibrary{fadings}

\definecolor{juliablue}{rgb}{0.251, 0.388, 0.847} 
\definecolor{juliagreen}{rgb}{0.22, 0.596, 0.149} 
\definecolor{juliapurple}{rgb}{0.584, 0.345, 0.698} 
\definecolor{juliared}{rgb}{0.796, 0.235, 0.2} 

\graphicspath{{images/}}

\usepackage{hyperref}

\usepackage[english]{babel}
\usepackage{hyphenat}
\usepackage{amsmath,amssymb,amsfonts,mathrsfs,latexsym,stmaryrd,mathtools}
\newcommand{\itadata}{\footnotesize \textsl{ITADATA2024: The 3$^{\text{rd}}$ Italian Conference on Big Data and Data Science}}
\usepackage{fancyhdr}
\fancypagestyle{empty}{%
  \fancyhf{}
  \fancyhead[C]{\itadata}
  \fancyhead[R]{\includegraphics[width=.05\textwidth]{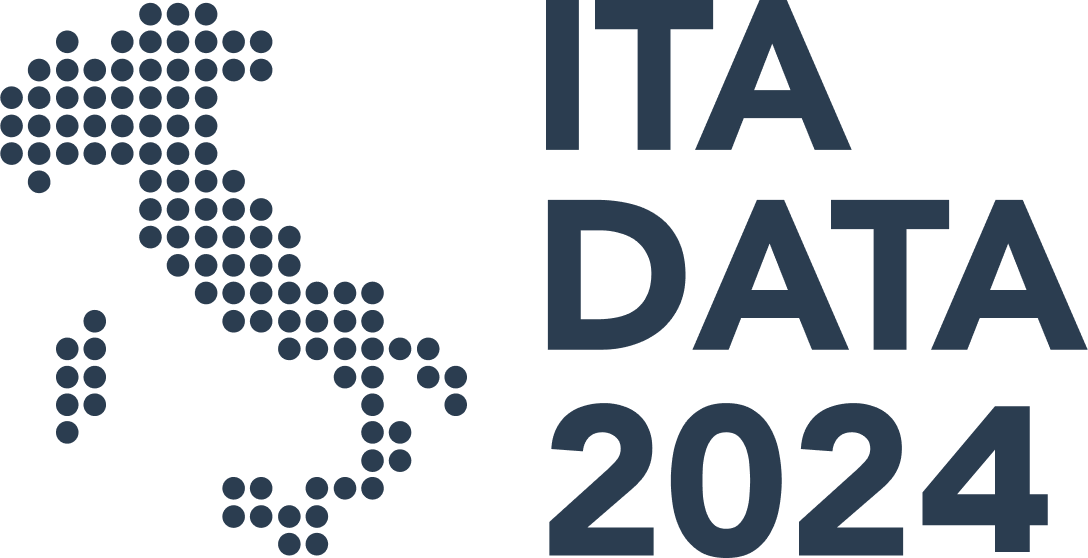}}
}

\newcommand{\exmodop}[1]{\ensuremath{\langle #1 \rangle}}

\newcommand{\univmodop}[1]{\ensuremath{[#1]}}

\begin{document}
\title{Symbolic Audio Classification via Modal Decision Tree Learning}
%
%
\author{Enrico Marzano\inst{1}\orcidID{0009-0001-1469-6497} \and\\
Giovanni Pagliarini\inst{2}\orcidID{0000-0002-8403-3250} \and\\
Riccardo Pasini\inst{2}\orcidID{0009-0009-2807-2245} \and\\
Guido Sciavicco\inst{2}\orcidID{0000-0002-9221-879X} \and\\
Ionel Eduard Stan\inst{3}\orcidID{0000-0001-9260-102X}}
\authorrunning{E. Marzano et al.}


%
\institute{
R\&D Deparment, Gap S.r.l.u.\\ Udine, Italy
\and
Applied Computational Logic and Artificial Intelligence (ACLAI) Lab,\\
Department of Mathematics and Computer Science, University of Ferrara\\
Ferrara, Italy\\
\and
Database Systems Group,
Faculty of Engineering, Free University of Bozen-Bolzano\\
Bolzano, Italy
}
\maketitle              
\begin{abstract}
The range of potential applications of acoustic analysis is wide. Classification of sounds, in particular, is a typical machine learning task that received a lot of attention in recent years. The most common approaches to sound classification are subsymbolic, typically based on neural networks,
and result in black-box models with high performances but very low transparency. In this work, we consider several audio tasks, namely, age and gender recognition, emotion classification, and respiratory disease diagnosis, and we approach them with a symbolic technique, that is, (modal) decision tree learning. We prove that such tasks can be solved using the same symbolic pipeline, that allows to extract simple rules with very high accuracy and low complexity. In principle, all such tasks could be associated to an autonomous conversation system, which could be useful in different contexts, such as an automatic reservation agent for an hospital or a clinic.

\end{abstract}
\section{Introduction}

Acoustic signal analysis represents one of the forefront challenges in modern Artificial Intelligence. This field spans various contexts and applications, such as speech recognition, speech identification, speech verification, conversational emotion and sentiment analysis, language recognition, medical diagnosis support, and music genre classification.

Recent years have seen a rise in applications to the real-time call monitoring scenario within call centers. In this context, automated telephonic systems utilizing intelligent audio analysis systems enable supervisors to track all calls handled by agents, alerting them promptly to any call events necessitating human intervention.
Several opportunities to elaborate on this framework exist: from systems for basic demographic detection (e.g., gender and age recognition from voice), to monitoring tools that notify operators of required interventions, up to sophisticated intelligent voice assistance systems, capable of fully serving the caller's needs akin to a human operator.
Towards a comprehensive assistance system, in this paper, we address several diverse speech-related problems, that is: gender and age recognition of the speaker; sentiment/emotion recognition of the speaker, and detection of respiratory diseases. Each of these problems is formulated as either a binary or multi-class classification task.


Modern approaches to audio classification predominantly rely on the so-called Mel-Frequency Cepstral Coefficients (MFCC)~\cite{Davis1980} feature extraction and subsymbolic model training, such as neural networks or support vector machines. These workflows often result in black-box systems that lack interpretability, thereby hindering crucial human involvement in the problem-solving process.
Contrary to this trend, in this work we approach the mentioned tasks using a symbolic approach. More in particular, we employ and test a comprehensive, open-source symbolic learning system known as SOLE~\cite{Solejl.gh} to solve such diverse set of acoustic learning tasks with the aim of finding, for each one them, a convenient trade-off between the performances of the obtained model and the simplicity of the logical rules that form them. SOLE is a open-source framework, written in the Julia programming language, that allows one to design and deploy end-to-end learning tasks, from data preprocessing and cleaning, filter-based feature extraction and selection, symbolic learning models training, testing, inspection, and post-hoc modification, and result visualization. The feature extraction and selection module, in particular, is able to connect to external packages, such as those specifically designed for acoustic features, and provides a workbench for the identification of the most informative ones.

The purpose of this work is to establish what performance level can rule-based symbolic models reach in acoustic recognition tasks, and examine and discuss the extracted rules. Moreover, a distinctive characteristics of the system SOLE is that it is able to extract classic symbolic models  based on propositional logic as well as modal and mixed propositional/modal logic symbolic models (see, e.g.,~\cite{DBLP:conf/aiia/MonicaPSS22,DBLP:conf/ecai/ManzellaPSS23,MANZELLA2023102486}) alike. Therefore, as a byproduct, we will be able to study the nature of the logical knowledge that is needed for each specific task. As a matter of fact, in the case of acoustic tasks the first level of approximation consists of simply extracting features from the entire sample (therefore reducing each sample to a numeric vector); with a modal symbolic learning system, instead, one can extract the same features from different temporal interval of each sample, and then study their temporal relationship. As a result, in some cases, it may happen that a classification rule for a given task emerges with the form, for example, {\em the average value of the feature $F_1$ is above a certain threshold while the minimum value of the feature $F_2$ is below a certain, different, threshold}, and such rule behaves better than a simpler one with the form, for instance, {\em the average value of the feature $F_1$ is above a certain threshold and the minimum value of the feature $F_2$ is below a certain, different, threshold}. As we shall see, it so happens that certain acoustic tasks are better solved with pure propositional rules, while other require the superior level of expressivity that modal rules allow, and establishing which is the case for each task is part of our objectives.



This work is organized as follows. In Section~\ref{sec:julia} we briefly describe the SOLE symbolic framework, and (audio) feature packages that are relevant to this study. Then, in Section~\ref{sec:results} we describe our experiments, including the datasets, experimental setting, and comment on the extracted symbolic rules, before concluding.

\section{Symbolic Learning from Audio Signals in Julia}~\label{sec:julia}

\noindent{\bf SOLE: a brief description.} SOLE~\cite{Solejl.gh} is a comprehensive framework for learning symbolic, logic-based models to extract knowledge from tabular and non-tabular data. As it can be seen in its high-level schema, shown in Fig.~\ref{fig1}, it includes three {\em core} packages, namely, {\em SoleData}, {\em SoleLogics}, and {\em SoleModels}, plus a set of {\em non-core} packages with different purposes.

\input{sole.tex}


The three core packages are focused on three different aspects of symbolic learning. {\em SoleData} contains tools to transform raw data into a form that is amenable to be used for logical learning. In the case of tabular data, a {\em dataset} is a finite collection of $m$ {\em instances} $\mathcal I=\{I_1,\ldots,I_m\}$, each described by the value of $n$ {\em attributes} $\mathcal A=\{A_1,\ldots,A_n\}$; if a tabular dataset is such that each instance is associated to a unique {\em label} from a set $\mathcal L=\{L_1,\ldots,L_k\}$, we call it {\em labelled tabular dataset}. In the symbolic context, instances are seen as models of some logical formalism. Then, learning a classifier from a dataset requires extracting the logical property that defines each class, and classification itself is performed by checking the learned formulas on the instance(s). To help this interpretation of classification, one takes into consideration that in the symbolic framework datasets are naturally associated to a logical alphabet $\mathcal P$ of atomic statements (the {\em inductive bias}), from which formulas are built. While in some cases alphabets are the result of a suitable attribute selection and/or domain filtration, from a purely methodological point of view we can assume an alphabet of conditions
\[
\mathcal P=\{(A\bowtie\  a) \mid A\in\mathcal A,a\in dom(A), \bowtie\  \in \{<,\le,=,\ge,>\}\},
\]
\noindent where $dom(A)$ is the {\em domain} of $A$, that is, the set of values that $A$ takes in $\mathcal I$. Thus, in the tabular case, a learned model can be identified with a {\em propositional logic formula}, such as, for example, {\em $A_1>4$ and $A_2\le 12$}.

Non-tabular data have relational components; for example, multi-variate time series have temporal components. A {\em time series} is an observation interpreted over a linear order; observations can be {\em univariate} if there is only one measurement, or {\em multivariate} if there are more than one; observed data type can be numerical or categorical. Thus a {\em temporal dataset} is a finite collection of $m$ {\em instances} $\mathcal I=\{I_1,\ldots,I_m\}$, each described $n$ {\em attributes} $\mathcal A=\{A_1,\ldots,A_n\}$ each one of them encompassing $N$ temporally ordered values; temporal datasets too can be labelled. The inductive bias for learning from a temporal dataset can take many forms. In the most intuitive case, one can always compare the value of an attribute $A$ with a threshold (as in the tabular case) {\em at a certain moment of time}; in this case, it would make sense to extract a classifier that has the form of a {\em point-based temporal logic formula}, such as, for instance, {\em $A_1>4$ and sometimes in the future $A_2<12$.} More generally, however, one can take advantage of some predetermined set of {\em feature extraction functions} $\mathcal F=\{f_1,\ldots,f_l\}$. Typically feature extraction functions for temporal data are functions that are applied to an entire time series to reduce it to a single number, but the same function can always be applied to an {\em interval} (even a instantaneous one) of the time series; then we can study a time series not only by the value of the function(s) applied to it, but by their relative qualitative positions. As a consequence, in its general form the set of atomic statements becomes
\[
\mathcal P=\{(f(A)\bowtie\  a) \mid A\in\mathcal A,a\in dom(f(A)), \bowtie\  \in \{<,\le,=,\ge,>\}, f \in \mathcal F\},
\]
\noindent and, in the non-tabular, specifically, temporal case, a learned model can be identified with a {\em propositional interval temporal logic formula} such as 
\[\exmodop{G}(\mathop{avg}(A_1)>4 \wedge (\univmodop{A} (\mathop{max}(A_2)\le 12) \vee \exmodop{O} (\mathop{min}(A_3)\le 20))),\]
which states that {\em there exists globally (G) an interval in which the average of $A_1$ is greater than $4$, that is either immediately followed by (A) all intervals in which the maximum of $A_2$ is less or equal than $12$, or overlapped by (O) an interval in which the maximum of $A_4$ is less or equal than $20$.}
Note that the set of relevant relations (e.g., ${G}$, ${A}$, ${O}$) is a parameter of the language.


{\em SoleLogics} collects the description of the logical languages that can be used on both sides, that is, raw datasets transformation into logisets and symbolic models learning. The SOLE framework is specifically designed to learn from temporal and spatial data, although the techniques and methods can be used to learn from other types of non-tabular data. Some noteworthy languages
are {\em Halpern and Shoham's Interval Temporal Logic} (HS)~\cite{DBLP:journals/jacm/HalpernS91} and its coarser fragments (e.g., HS3 and HS7)~\cite{DBLP:journals/ai/Munoz-VelascoPS19} for the temporal case, and {\em Lutz and Wolter's Modal Logic of Topological Relations} RCC8, and its coarser fragments (e.g., RCC5, RCC3)~\cite{LRCC8})
for the spatial case. These temporal and spatial logics are particular cases of the more general {\em modal} logic~\cite{blackburn_rijke_venema_2001}, which is archetypical of a class of more-than-propositional logics with high expressive power and (sometimes) low computational complexity for deduction.


{\em SoleModels} provides general definitions for symbolic models, which are produced by learning algorithms, implemented in non-core packages.


Among the non-core packages, two are relevant to our discussion. The first one is {\em (modal) decision trees}, that contains the definitions and the learning algorithms for tree-based classification models for learning decision tree and forests in propositional and modal logics. Modal decision trees, which can be declined in temporal and spatial decision trees, were introduced in~\cite{DBLP:conf/jelia/BrunelloSS19,DBLP:conf/time/SciaviccoS20,DBLP:conf/aiia/MonicaPSS22,DBLP:conf/aiia/ManzellaPSS23} and already used for acoustic analysis, for example, in~\cite{MANZELLA2023102486}. The second one is {\em SoleFeatures}~\cite{DBLP:conf/itadata/CavinaMPSS23}, that contains the necessary tools for using feature extraction functions as above described. Such functions can be borrowed from existing packages or built for a specific exercise; in the case of learning from acoustic data, the Julia package {\em Audio911} is a possible source of several well-known audio features.


\noindent{\bf Experimental flow and parameters.} Specifically to the treatment of acoustic data, we can summarize our proposed operational flow as follows.


A raw audio sample consists of a single channel audio signal. Such a signal is, first, processed via a set of audio features that typically include: the 12 among the classical spectral audio features~\cite{giannakopoulos2014introduction}, 26 features derived from the {\em Mel spectrogram}~\cite{stevens1937scale} of the signal, and 39 derived from the {\em MFCC} transform~\cite{Davis1980}, for a total of 77 features. The end result consists, therefore, of describing an audio signal as a multivariate time series with 77 attributes, which is obtained as follows. The first step consists of computing the linear spectrogram, from which the typical audio features such as F0 (fundamental frequency), the {\em centroid}, the {\em crest}, and so on, can be immediately computed. The spectrogram is, then, converted and represented in the so-called {\em Mel scale}, which is a logarithmic representation which causes the frequency space to better reflect human ear perception of frequencies, and the set of $26$ triangular band-pass filters is convolved across the frequency spectrum, discretizing it into a finite number of frequencies. The resulting discretized spectrogram reports the time-distributed sound power for $26$ specific frequencies evenly spaced across a logarithmic axis.
Finally, a discrete cosine transform  is applied to the logarithm of the discretized spectrogram along the frequency axis, which compresses the spectral information at each point in time into a fixed number of coefficients (13), to each one of which we compute the first and the second derivative (commonly referred to as {\em Delta} and {\em Delta-Delta}).

The resulting multivariate time series may then undergo a second-level univariate feature extraction process. We fix the set $\mathcal F$ to contain the {\em maximum}, {\em minimum}, {\em average}, {\em median} and {\em standard deviation} functions, plus the set of so-called {\em symbolic Catch22 features}~\cite{DBLP:journals/datamine/LubbaSKSFJ19}, that is, {\em entropy{\_}pairs}, {\em transition{\_}variance}, {\em stretch{\_}high}, and {\em stretch{\_}decrease}. Such features have an immediate interpretation; for example, {\em stretch{\_}decrease} computes the longest sequence of successive time points where the value of the attribute decreases. Then, as explained above, these functions can either be applied to entire series, ultimately producing a tabular dataset from which a standard decision tree can be learned, or can be used within a modal decision tree learning algorithm in a native way.
%


%
%
%
%
%
%

\section{Experiments and Results}~\label{sec:results}

\input{data.tex}


\input{experimental-setting.tex}


\input{discussion.tex}

\section{Conclusion}

In this work, we have deployed the SOLE open-source, symbolic learning framework on modern learning problems involving voice samples. We have tackled different tasks using the same symbolic learning method and the same parametrization, ultimately obtaining simple classification models whose performances are comparable to those of hyper-parametrized, subsymbolic, and usually computationally challenging systems. The extracted rules are ready for interpretation, and allow for further inspection and discussion.

Acoustic analysis has several potential applications. Integrated models that can, in fact, solve very different problems from the same voice samples can be applied in many contexts; one example is automated call centers, which in some cases assist customer care in hospitals, clinics, and health centers: ideally, it is possible to think of integrated systems that extract a number of useful information on the caller, real-time, with a high degree of accuracy, to ease a precise and goal-centered care.



\begin{credits}

\vspace{-0.3cm}

\subsubsection{\ackname} This research has been partially funded by the FIRD project {\em Modal Geometric Symbolic Learning} (University of Ferrara), the project {\em Conversational Automation and Voice\&Speech Analysis for Active and Assisted Living}, funded by the Italian Region Friuli Venezia Giulia, and the Gruppo Nazionale Calcolo Scientifico-Istituto Nazionale di Alta Matematica (INDAM-GNCS) project {\em Symbolic and Numerical Analysis of Cyberphysical Systems}, CUP code E53C22001930001. Giovanni Pagliarini, Guido Sciavicco, and Ionel Eduard Stan are GNCS-INdAM members.

\vspace{-0.3cm}

\subsubsection{\discintname}

The following private company won the public call through which the project {\em Conversational Automation and Voice\&Speech Analysis for Active and Assisted Living} was funded: GAP S.r.l.u. Enrico Marzano is CIO of GAP AUTOMATION, the R\&D unit of GAP S.r.l.u.
\end{credits}

\bibliographystyle{splncs04}
\bibliography{biblio}

\end{document}

%% file: sole.tex
\colorlet{solesymboliclogic}{juliapurple!80!white}
\colorlet{solesymbolicdata}{juliablue!80!white}
\colorlet{solesymbolicmodels}{juliagreen!80!white}
\colorlet{solesymboliclearning}{juliared!80!white}
\begin{figure}[tp]
\centering
		\resizebox{0.8\textwidth}{!}{%
			\begin{tikzpicture}[xscale=1.2,text centered, every text node part/.style={align=center,font=\small},>=Stealth]
				\tikzstyle{solenode}=[rectangle,rounded corners,inner sep=0.14cm,draw=black,font=\normalsize]

				\node[solenode,fill=solesymboliclogic] (SL) at (0,-0.50) {SoleLogics.jl};

				\node[solenode,fill=solesymbolicdata] (SD) at (-1.2,0.50) {SoleData.jl};
				\node[solenode,fill=solesymbolicmodels] (SM) at (1.2,0.50) {SoleModels.jl};

				\draw[->] (SM) -- (SL);
				\draw[->] (SD) -- (SL);
				{
				}
				{

				}

				\node[rectangle,draw=black!50,dashed,minimum height=2.4cm,minimum width=5cm,xshift=0.08cm] (SOLE) at (0,0) {};

			\begin{scope}[yshift=1.4cm]
				\node at (0,4.5) {
					\includegraphics[scale=0.3]{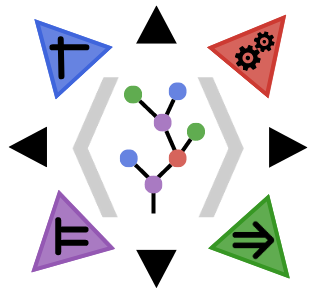}
				};
				\node at (0,3) {Sole.jl};

			\end{scope}


{
				\node[solenode,fill=solesymbolicdata] (SF) at (-3.8,1.2) {SoleFeatures.jl};
				\draw[->] (SF) -- (SOLE);

				\node[solenode,fill=solesymbolicdata] (SV) at (-3.8,0.2) {SoleViz.jl};
				\draw[->] (SV) -- (SOLE);
}


{
				\node[solenode,fill=solesymboliclearning] (SEX) at (0,3.6) {SoleXperimenter.jl}; \draw[->] (SEX) -- (SOLE);
				\node[solenode,fill=solesymboliclearning] (MDL) at (-1.8,2.9) {{Modal}DecisionLists.jl}; \draw[->] (MDL) -- (SOLE);
				\node[solenode,fill=solesymboliclearning] (MIF) at (1.8,2.9) {{Modal}IsolationForests.jl}; \draw[->] (MIF) -- (SOLE);
				\node[solenode,fill=solesymboliclearning] (MDT) at (-2.8,2.2) {{Modal}DecisionTrees.jl}; \draw[->] (MDT) -- (SOLE);
				\node[solenode,fill=solesymboliclearning] (MAR) at (2.8,2.2) {{Modal}AssociationRules.jl}; \draw[->] (MAR) -- (SOLE);

}
{
				\node[solenode,fill=solesymbolicmodels] (SPH) at (3.8,1.2) {SolePostHoc.jl};
				\draw[->] (SPH) -- (SOLE);

				\node[solenode,fill=solesymbolicmodels] (SLR) at (3.8,0.2) {SoleReasoners.jl};
				\draw[->] (SLR) -- (SOLE);
}
			\end{tikzpicture}
}
\caption{Structure of the Sole.jl framework for symbolic AI. Packages in green provide tools for manipulating logical formulas; packages in red provide tools for (symbolic) data processing; those in blue provide tools for learning symbolic models; those in purple provide tools for manipulating (symbolic) models.} \label{fig1}
\end{figure}
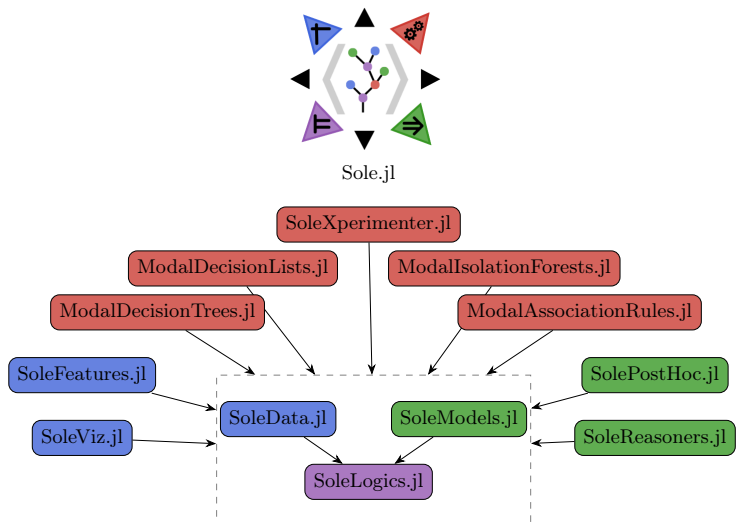

%% file: data.tex
{
\begin{table}[tp]
\caption{Description of the 11 audio classification tasks designed for this work.}\label{datasets}
\centering
\normalsize
\resizebox{0.98\linewidth}{!}{
\begin{tabular}{|l|l|l|r|r|r|}
\hline
Task type &
Task	&
Class names	&
\# classes &
\# inst./class &
audio length
\\
\hline
\multirow{1}{*}{{Gender (G)}} & G2 & m/f &	2	&	1000	&	6.00s
\\	\hline
\multirow{2}{*}{{Age (A)}}	&	A2 & [0--39]/60+ &	2	&	1000	&	6.00s
\\		&	A4 & [0--29]/[30--49]/[50--69]/70+ &	4	&	1000	&	6.00s
\\	\hline
\multirow{2}{*}{{Emotion (E)}}		&	E2 & positive/negative &	2	&	672	&	1.92s
\\		&	E8 & neu/cal/hap/sur/sad/ang/fea/dis &	8	&	384	&	1.92s
\\	\hline
\multirow{5}{*}{{Respiratory (R)}}			&	R2a & healthy/bronchiectasis &	2	&	88	&	2.25s
\\		&	R2b & healthy/bronchiolitis &	2	&	106	&	1.30s
\\		&	R2c & healthy/COPD &	2	&	132	&	2.18s
\\		&	R2d & healthy/pneumonia &	2	&	132	&	2.18s
\\		&	R2e & healthy/URTI &	2	&	114	&	1.75s
\\		&	R4 & healthy/COPD/pneumonia/URTI &	4	&	203	&	2.11s
\\
\hline
\end{tabular}
}
\end{table}
}

\noindent{\bf Experimental design.} We designed 11 tasks, including a task of
gender recognition ({G}), two tasks of
age recognition ({A}), two tasks of
emotion recognition ({E}), and six tasks of
respiratory disease detection ({R}). The resulting classification datasets for the 11 tasks are shown in Tab.~\ref{datasets}.

\smallskip

For gender and age recognition tasks, we used speech samples in Chinese, English, French, German, Italian, Spanish, and Swedish available in the Mozilla Common Voice Corpus 17.0 dataset~\cite{ardila2019common}. All audio samples were provided with validated age and gender labels.
Each audio sample was stripped of parts without speech. To this purpose, a simple speech detector was used, removing parts of the audio where the overall signal strength was irregular or not present above a certain threshold.
%
%
Ultimately, this procedure led to a pool of 29415 samples audio samples of at least 6 seconds of speech;
all samples were, then, reduced to the smallest length.
As no sample was available from transgender or non-binary subjects, the only gender recognition task (G2) we designed is a binary classification task.
The age labels, instead, were already discretized into teens (age < 19), twienties (age from 19 to 29), etc., up to nineties (age $\geq$ 89). While, originally, this task is a regression one, we treated it as classification with 2 and 4 classes, which are derived by aggregation of continuous bins. In accordance with the number of samples available, and their class-balance, we opted for a [0--39]/(60+) binary binning, and a [0--29]/[30--49]/[50--69]/70+ quaternary binning.

For emotion recognition tasks, audio recordings of actors portraying various emotions from the RAVDESS emotional speech audio dataset~\cite{RAVDESS} were used. The dataset was originally collected and tailored specifically for this purpose, so the samples are already edited and normalized, and no preprocessing was necessary. Audios in the dataset are divided into 8 categories: calm, happy, sad, angry, fearful, surprise, and disgust expressions.
In addition to classifying the 8 emotions, we designed a coarser task for analyzing the differences between calm, happy and surprise expressions that can, arguably, be classified as positive and sad, angry, fearful, and disgust, those that can be classified as negative.

For respiratory disease detection, the Respiratory Sound Database~\cite{rocha2018alpha} was used. The dataset includes of recordings of breaths captured using specific microphones, mostly contact microphones. Consequently, audio recordings are often compromised by significant mechanical noise due to the natural application of the microphone. We isolated these noises via a band-pass filter from 300Hz to 4000Hz, filtering out frequencies {that are not informative for breath analysis} and where the highest amplitude of mechanical noise is concentrated. The signal is extremely dynamic, with significant discrepancies between low-volume and high-volume parts. Therefore, it is not possible to apply any kind of noise gate to eliminate the low-volume parts; rather, it is deemed appropriate to analyze the audio in its entirety for future studies. Given the delicacy of the samples to be analyzed, a binary approach was preferred, evaluating the differences between healthy individuals and those with disorders such as: bronchiectasis, bronchiolitis, Chronic Obstructive Pulmonary Disease (COPD), pneumonia, and Upper Respiratory Tract Infection (URTI).
With the aim of expanding the analysis, we also designed a quaternary classification task involving healthy, COPD, pneumonia and URTI individuals.

\smallskip

The audio preprocessing was performed using the Julia package {\em Audio911} for audio feature extraction.
After a pre-screening step, where different parametrizations were tested, all audios were ultimately resampled at a rate of 8000Hz,
and the linear spectrogram was computed via a Short-Time Fourier Transform (STFT) was applied to the samples, based on Hanning windows of length of 256 samples, with overlap between contiguous windows of 128 samples.
From the spectrogram, the 77-attributes multivariate time series was computed.
Finally, the resulting time series were downsized via a moving average aggregation filter, reducing the number of temporal points to 5, via average windows with a $20\%$ overlap between contiguous windows.



%% file: experimental-setting.tex
For each of the 11 tasks, decision trees (DTs) and decision forests (DFs) models were trained, in both propositional and temporal versions, via the modal decision tree and modal random forest implementations provided in the {\em ModalDecisionTrees} package~\cite{ModalDecisionTrees}. Temporal trees were trained based on the HS7 interval temporal logic~\cite{DBLP:journals/ai/Munoz-VelascoPS19}, which includes, among other relations,
{\em later} ($\mathop{L}$),
{\em later inverse} ($\mathop{\overline{L}}$),
{\em proper part} ($\mathop{DBE}$), and
{\em proper part inverse} ($\mathop{\overline{DBE}}$). To minimize the bias, datasets were balanced by downsampling the majority class and randomly split into two (balanced) sets for training (80\%) and test (20\%). This process was repeated $10$ times (via randomized 10-folds cross-validation), and the average and standard deviation of the performance metrics were considered. As for the parametrization of random forests, after a pre-screening phase, we set each tree to be trained on the $70\%$ of the cardinality of each training set ($m$) and $50\%$ of the number of attributes, and we learned 100 trees in each round.  While experiments for single trees were run with a standard pre-pruning setting, that is, minimum entropy gain of $0.01$ for a split to be meaningful and maximum entropy at each leaf of $0.6$, forests grow full trees. In all cases we let $\bowtie$ be in $\{\le,\ge\}$, and used the full set $\mathcal F$ with 9 second-level feature extraction functions.
Experiments were run on a computing server with \emph{Intel\textsuperscript{\textregistered} Xeon\textsuperscript{\textregistered} Gold 6238R} using $30$ threads, and the total run time was of $16.5$ hours. Code for replicating the proposed pipeline is available at \url{https://github.com/aclai-lab/audio-rules2024}.




%% file: discussion.tex
\input{table.tex}

\noindent{\bf Results.}
Tab.~\ref{tab1} reports metrics of performance and symbolic complexity measured in cross-validation .
A first glance reveals that the method is competitive in absolute terms. In the most difficult binary task (namely, E2, classifying between positive and negative emotions), the best performing model (in this case, modal forest) attains a 70.9\% accuracy analyzing, we recall, audio samples of length 1.92 seconds. In the easiest task (which turned out to be R2a, distinguishing between a healthy subject and a subject with bronchiectasis), the best model (again, modal forest) returns a 99.3\% accuracy, with as little as 1\% of standard deviation. Similarly, in the the most difficult non-binary classification task (A4, classifying among four age groups) propositional forests reached 68.9\% accuracy (to be compared with the 25\% accuracy that would be expected from a random classifier), while in the easiest one (R4, distinguishing between healthy subjects, subjects with COPD, subjects with pneumonia, and subjects with URTI), propositional random forests attain up to 81.5\% accuracy. Comparing our results with existing results in the literature, we observe that while for some of the tasks, such as gender recognition, existing models can be quite accurate (e.g., 99.6\% in~\cite{kwasny2021gender}), in other cases, such as age classification, our solution seems slightly more performing (e.g., compared with~\cite{zaman2021one}, in which an accuracy of 70.8\% is declared). Similar considerations concerning other tasks can be made.

These results, however, should be analyzed in the light of the simplicity of the learned models. Consider the case of gender recognition, for example. The best modal forest for this task shows 94.7\% accuracy with 57.5 rules per tree, on average; accepting a 6\% accuracy drop, however, one could perform binary gender recognition with as little as two propositional rules. Another example of simple models is bronchiectasis recognition, where, less than two (propositional or modal) rules are needed to reach over 99\% of accuracy in cross-validation.


\input{rules.tex}

In a few cases, in terms of the result obtained with forests, the increased expressive power that temporal reasoning allows over the propositional approach seems relevant. These are: the gender recognition task (with a little increase of 0.2\% accuracy), the emotion recognition tasks (0.4\% increase), and four of out six respiratory disease tasks (up to 1.6\% increase). In several cases, moreover, the number of rules that are needed in the temporal case decreases, compared with the propositional one; a representative example is that of R2d (healthy subjects vs. subjects with pneumonia), where modal decision trees need, on average, 11.5 rules, against 19.6 rules for the propositional one.


As a final analysis, we inspect the classification rules enclosed in the trained tree models. For the purpose, we consider a subset of binary classification tasks, train 3 propositional and 3 modal trees in a 80\%/20\% balanced setting, and extract all rules with a test confidence greater than $0.50,$ and a number of covered test instances (or {\em coverage}) greater than 8.
Tab.~\ref{rules} shows all such rules, extracted from tree models for tasks E2, R2a, R2b, R2c, R2d, and R2e; recall that the number of test instances for these tasks are 402, 54, 63, 78, 78, and 69, respectively.
As it can be seen, some well-performing propositional and modal rules exist for each of these tasks, with rule confidences estimated between $0.66$ and $1.00$. This result indicates that trees with suboptimal performance can still encompass rules with optimal performances on parts of the classification task.
Furthermore, some well-performing propositional and modals rules stand out for they symbolic simplicity. 
Of particular interest is the case of R2a (bronchiectasis classification), where all well-performing rules score a 100\% test confidence, cover at least 15 test instances, and have reduced a symbolic complexity. For example, the first propositional rule can be read as ``IF, across the series, the standard deviation of the second Mel-Frequency Cepstral Coefficient is greater than 2.46, THEN the patient is sick''. Similarly the last modal rule can be read as ``IF, across the series, there exists an interval where the spectral feature DECREASE is never greater than -4.55, and frequency 353Hz is never louder than 0.01, THEN the patient is healthy''.
%
%
%
In the case of other tasks, well-performing rules are more complex.
%
Altogether, the results indicate that suboptimal trees can still encompass interpretable rules, providing both optimal performance and useful insights on parts of the classification task.

%% file: table.tex
{
\begin{table}[t!]
\caption{Cross-validation results
for
tree-based models learned on the 11 tasks. For each dataset and method, the average and standard deviation of the $\kappa$ coefficient, the overall accuracy ($\alpha$) and the number of tree leaves ($\rho_{\tau}$) is reported. For each line, the highest value for $\alpha$ and the smallest value for $\rho_{\tau}$ are highlighted.}\label{tab1}
\centering
\normalsize
\resizebox{.98\linewidth}{0.28\textheight}{
\begin{tabular}{|c|ll||*{2}{r|rr|}}
\hline
	&	&	&	\multicolumn{6}{|c||}{Decision Tree}
\\
	&	&	&	\multicolumn{3}{c|}{Propositional}
	&	\multicolumn{3}{c||}{Modal}
\\
	&	\multicolumn{2}{c||}{Task}	&
\multicolumn{1}{c|}{$\kappa$}	&
\multicolumn{1}{c}{$\alpha$}	&
\multicolumn{1}{c|}{$\rho_{\tau}$}	&
\multicolumn{1}{c|}{$\kappa$}	&
\multicolumn{1}{c}{$\alpha$}	&
\multicolumn{1}{c||}{$\rho_{\tau}$}
\\
\hline
\multirow{1}{*}{{G}}		&	G2	&	&			
${77.2} \pm \pgfmathprintnumber[fixed,fixed zerofill,precision=0]{1.52}$	&	${88.6} \pm \pgfmathprintnumber[fixed,fixed zerofill,precision=0]{0.76}$	&	${2.0} \pm \pgfmathprintnumber[fixed,fixed zerofill,precision=0]{0.00}$
&	${77.2} \pm \pgfmathprintnumber[fixed,fixed zerofill,precision=0]{1.52}$	&	${88.6} \pm \pgfmathprintnumber[fixed,fixed zerofill,precision=0]{0.76}$	&	${2.0} \pm \pgfmathprintnumber[fixed,fixed zerofill,precision=0]{0.00}$
\\	\hline
\multirow{2}{*}{{A}}		&	A2	&	&			
${31.4} \pm \pgfmathprintnumber[fixed,fixed zerofill,precision=0]{3.11}$	&	${65.7} \pm \pgfmathprintnumber[fixed,fixed zerofill,precision=0]{1.55}$	&	$\mathbf{59.9} \pm \pgfmathprintnumber[fixed,fixed zerofill,precision=0]{36.16}$
&	${31.6} \pm \pgfmathprintnumber[fixed,fixed zerofill,precision=0]{3.53}$	&	$\mathbf{65.8} \pm \pgfmathprintnumber[fixed,fixed zerofill,precision=0]{1.76}$	&	${70.0} \pm \pgfmathprintnumber[fixed,fixed zerofill,precision=0]{30.37}$
\\											&	A4	&	&			
${37.6} \pm \pgfmathprintnumber[fixed,fixed zerofill,precision=0]{2.51}$	&	$\mathbf{53.2} \pm \pgfmathprintnumber[fixed,fixed zerofill,precision=0]{1.88}$	&	${305.2} \pm \pgfmathprintnumber[fixed,fixed zerofill,precision=0]{41.41}$
&	${37.3} \pm \pgfmathprintnumber[fixed,fixed zerofill,precision=0]{2.87}$	&	${52.9} \pm \pgfmathprintnumber[fixed,fixed zerofill,precision=0]{2.15}$	&	$\mathbf{255.1} \pm \pgfmathprintnumber[fixed,fixed zerofill,precision=0]{54.84}$
\\	\hline
\multirow{2}{*}{{E}}		&	E2	&	&			
${27.4} \pm \pgfmathprintnumber[fixed,fixed zerofill,precision=0]{5.39}$	&	$\mathbf{63.7} \pm \pgfmathprintnumber[fixed,fixed zerofill,precision=0]{2.70}$	&	${52.8} \pm \pgfmathprintnumber[fixed,fixed zerofill,precision=0]{15.48}$
&	${26.7} \pm \pgfmathprintnumber[fixed,fixed zerofill,precision=0]{4.52}$	&	${63.4} \pm \pgfmathprintnumber[fixed,fixed zerofill,precision=0]{2.26}$	&	$\mathbf{44.2} \pm \pgfmathprintnumber[fixed,fixed zerofill,precision=0]{18.89}$
\\											&	E8	&	&			
${15.6} \pm \pgfmathprintnumber[fixed,fixed zerofill,precision=0]{3.10}$	&	$\mathbf{26.1} \pm \pgfmathprintnumber[fixed,fixed zerofill,precision=0]{2.72}$	&	$\mathbf{102.6} \pm \pgfmathprintnumber[fixed,fixed zerofill,precision=0]{21.73}$
&	${14.3} \pm \pgfmathprintnumber[fixed,fixed zerofill,precision=0]{2.31}$	&	${25.0} \pm \pgfmathprintnumber[fixed,fixed zerofill,precision=0]{2.02}$	&	${107.4} \pm \pgfmathprintnumber[fixed,fixed zerofill,precision=0]{5.62}$
\\	\hline
\multirow{5}{*}{{R}}		&	R2a	&	&			
${96.7} \pm \pgfmathprintnumber[fixed,fixed zerofill,precision=0]{3.89}$	&	${98.3} \pm \pgfmathprintnumber[fixed,fixed zerofill,precision=0]{1.94}$	&	${2.0} \pm \pgfmathprintnumber[fixed,fixed zerofill,precision=0]{0.00}$
&	${98.3} \pm \pgfmathprintnumber[fixed,fixed zerofill,precision=0]{2.69}$	&	$\mathbf{99.2} \pm \pgfmathprintnumber[fixed,fixed zerofill,precision=0]{1.34}$	&	${2.0} \pm \pgfmathprintnumber[fixed,fixed zerofill,precision=0]{0.00}$
\\											&	R2b	&	&			
${61.8} \pm \pgfmathprintnumber[fixed,fixed zerofill,precision=0]{8.94}$	&	$\mathbf{80.9} \pm \pgfmathprintnumber[fixed,fixed zerofill,precision=0]{4.47}$	&	${9.4} \pm \pgfmathprintnumber[fixed,fixed zerofill,precision=0]{1.35}$
&	${58.2} \pm \pgfmathprintnumber[fixed,fixed zerofill,precision=0]{6.79}$	&	${79.1} \pm \pgfmathprintnumber[fixed,fixed zerofill,precision=0]{3.39}$	&	$\mathbf{8.6} \pm \pgfmathprintnumber[fixed,fixed zerofill,precision=0]{1.90}$
\\											&	R2c	&	&			
${79.6} \pm \pgfmathprintnumber[fixed,fixed zerofill,precision=0]{5.34}$	&	${89.8} \pm \pgfmathprintnumber[fixed,fixed zerofill,precision=0]{2.67}$	&	$\mathbf{3.3} \pm \pgfmathprintnumber[fixed,fixed zerofill,precision=0]{0.95}$
&	${80.6} \pm \pgfmathprintnumber[fixed,fixed zerofill,precision=0]{4.85}$	&	$\mathbf{90.3} \pm \pgfmathprintnumber[fixed,fixed zerofill,precision=0]{2.43}$	&	${7.4} \pm \pgfmathprintnumber[fixed,fixed zerofill,precision=0]{1.84}$
\\											&	R2d	&	&			
${75.5} \pm \pgfmathprintnumber[fixed,fixed zerofill,precision=0]{6.28}$	&	$\mathbf{87.7} \pm \pgfmathprintnumber[fixed,fixed zerofill,precision=0]{3.15}$	&	${19.6} \pm \pgfmathprintnumber[fixed,fixed zerofill,precision=0]{6.80}$
&	${74.7} \pm \pgfmathprintnumber[fixed,fixed zerofill,precision=0]{7.42}$	&	${87.4} \pm \pgfmathprintnumber[fixed,fixed zerofill,precision=0]{3.71}$	&	$\mathbf{11.5} \pm \pgfmathprintnumber[fixed,fixed zerofill,precision=0]{1.35}$
\\											&	R2e	&	&			
${37.0} \pm \pgfmathprintnumber[fixed,fixed zerofill,precision=0]{7.31}$	&	${68.5} \pm \pgfmathprintnumber[fixed,fixed zerofill,precision=0]{3.65}$	&	$\mathbf{16.1} \pm \pgfmathprintnumber[fixed,fixed zerofill,precision=0]{3.78}$
&	${38.8} \pm \pgfmathprintnumber[fixed,fixed zerofill,precision=0]{8.01}$	&	$\mathbf{69.4} \pm \pgfmathprintnumber[fixed,fixed zerofill,precision=0]{4.01}$	&	${30.8} \pm \pgfmathprintnumber[fixed,fixed zerofill,precision=0]{7.69}$
\\											&	R4	&	&			
${58.9} \pm \pgfmathprintnumber[fixed,fixed zerofill,precision=0]{2.92}$	&	${69.1} \pm \pgfmathprintnumber[fixed,fixed zerofill,precision=0]{2.19}$	&	${30.6} \pm \pgfmathprintnumber[fixed,fixed zerofill,precision=0]{8.47}$
&	${59.5} \pm \pgfmathprintnumber[fixed,fixed zerofill,precision=0]{4.37}$	&	$\mathbf{69.6} \pm \pgfmathprintnumber[fixed,fixed zerofill,precision=0]{3.28}$	&	$\mathbf{28.8} \pm \pgfmathprintnumber[fixed,fixed zerofill,precision=0]{7.38}$
\\
\hline	&	&	&	\multicolumn{6}{|c||}{Decision Forest}
\\
	&	&	&	\multicolumn{3}{c|}{Propositional}
	&	\multicolumn{3}{c||}{Modal}
\\
	&	\multicolumn{2}{c||}{Task}	&
\multicolumn{1}{c|}{$\kappa$}	&
\multicolumn{1}{c}{$\alpha$}	&
\multicolumn{1}{c|}{$\rho_{\tau}$}	&
\multicolumn{1}{c|}{$\kappa$}	&
\multicolumn{1}{c}{$\alpha$}	&
\multicolumn{1}{c||}{$\rho_{\tau}$}
\\
\hline
\multirow{1}{*}{{G}}		&	G2	&	&			
${89.1} \pm \pgfmathprintnumber[fixed,fixed zerofill,precision=0]{1.06}$	&	${94.5} \pm \pgfmathprintnumber[fixed,fixed zerofill,precision=0]{0.53}$	&	${58.0} \pm \pgfmathprintnumber[fixed,fixed zerofill,precision=0]{1.06}$
&	${89.4} \pm \pgfmathprintnumber[fixed,fixed zerofill,precision=0]{0.95}$	&	$\mathbf{94.7} \pm \pgfmathprintnumber[fixed,fixed zerofill,precision=0]{0.47}$	&	$\mathbf{57.5} \pm \pgfmathprintnumber[fixed,fixed zerofill,precision=0]{0.87}$
\\	\hline
\multirow{2}{*}{{A}}		&	A2	&	&			
${51.8} \pm \pgfmathprintnumber[fixed,fixed zerofill,precision=0]{2.27}$	&	$\mathbf{75.9} \pm \pgfmathprintnumber[fixed,fixed zerofill,precision=0]{1.13}$	&	${165.2} \pm \pgfmathprintnumber[fixed,fixed zerofill,precision=0]{0.66}$
&	${51.0} \pm \pgfmathprintnumber[fixed,fixed zerofill,precision=0]{1.93}$	&	${75.5} \pm \pgfmathprintnumber[fixed,fixed zerofill,precision=0]{0.97}$	&	$\mathbf{152.3} \pm \pgfmathprintnumber[fixed,fixed zerofill,precision=0]{0.70}$
\\											&	A4	&	&			
${59.8} \pm \pgfmathprintnumber[fixed,fixed zerofill,precision=0]{1.14}$	&	$\mathbf{69.8} \pm \pgfmathprintnumber[fixed,fixed zerofill,precision=0]{0.86}$	&	${547.4} \pm \pgfmathprintnumber[fixed,fixed zerofill,precision=0]{1.68}$
&	${59.6} \pm \pgfmathprintnumber[fixed,fixed zerofill,precision=0]{1.16}$	&	${69.7} \pm \pgfmathprintnumber[fixed,fixed zerofill,precision=0]{0.87}$	&	$\mathbf{510.4} \pm \pgfmathprintnumber[fixed,fixed zerofill,precision=0]{1.02}$
\\	\hline
\multirow{2}{*}{{E}}		&	E2	&	&			
${41.7} \pm \pgfmathprintnumber[fixed,fixed zerofill,precision=0]{3.13}$	&	${70.9} \pm \pgfmathprintnumber[fixed,fixed zerofill,precision=0]{1.56}$	&	${61.9} \pm \pgfmathprintnumber[fixed,fixed zerofill,precision=0]{0.29}$
&	${42.6} \pm \pgfmathprintnumber[fixed,fixed zerofill,precision=0]{2.89}$	&	$\mathbf{71.3} \pm \pgfmathprintnumber[fixed,fixed zerofill,precision=0]{1.44}$	&	$\mathbf{57.4} \pm \pgfmathprintnumber[fixed,fixed zerofill,precision=0]{0.32}$
\\											&	E8	&	&			
${32.4} \pm \pgfmathprintnumber[fixed,fixed zerofill,precision=0]{3.32}$	&	${40.8} \pm \pgfmathprintnumber[fixed,fixed zerofill,precision=0]{2.90}$	&	${93.0} \pm \pgfmathprintnumber[fixed,fixed zerofill,precision=0]{0.60}$
&	${32.9} \pm \pgfmathprintnumber[fixed,fixed zerofill,precision=0]{3.72}$	&	$\mathbf{41.3} \pm \pgfmathprintnumber[fixed,fixed zerofill,precision=0]{3.25}$	&	$\mathbf{89.1} \pm \pgfmathprintnumber[fixed,fixed zerofill,precision=0]{0.56}$
\\	\hline
\multirow{5}{*}{{R}}		&	R2a	&	&			
${95.6} \pm \pgfmathprintnumber[fixed,fixed zerofill,precision=0]{3.51}$	&	${97.8} \pm \pgfmathprintnumber[fixed,fixed zerofill,precision=0]{1.76}$	&	$\mathbf{1.5} \pm \pgfmathprintnumber[fixed,fixed zerofill,precision=0]{0.01}$
&	${98.7} \pm \pgfmathprintnumber[fixed,fixed zerofill,precision=0]{1.76}$	&	$\mathbf{99.3} \pm \pgfmathprintnumber[fixed,fixed zerofill,precision=0]{0.88}$	&	${1.6} \pm \pgfmathprintnumber[fixed,fixed zerofill,precision=0]{0.03}$
\\											&	R2b	&	&			
${71.4} \pm \pgfmathprintnumber[fixed,fixed zerofill,precision=0]{7.13}$	&	${85.7} \pm \pgfmathprintnumber[fixed,fixed zerofill,precision=0]{3.57}$	&	$\mathbf{10.1} \pm \pgfmathprintnumber[fixed,fixed zerofill,precision=0]{0.22}$
&	${73.8} \pm \pgfmathprintnumber[fixed,fixed zerofill,precision=0]{8.13}$	&	$\mathbf{86.9} \pm \pgfmathprintnumber[fixed,fixed zerofill,precision=0]{4.07}$	&	${10.3} \pm \pgfmathprintnumber[fixed,fixed zerofill,precision=0]{0.18}$
\\											&	R2c	&	&			
${90.8} \pm \pgfmathprintnumber[fixed,fixed zerofill,precision=0]{4.26}$	&	${95.4} \pm \pgfmathprintnumber[fixed,fixed zerofill,precision=0]{2.13}$	&	$\mathbf{11.4} \pm \pgfmathprintnumber[fixed,fixed zerofill,precision=0]{0.44}$
&	${91.2} \pm \pgfmathprintnumber[fixed,fixed zerofill,precision=0]{4.44}$	&	$\mathbf{95.6} \pm \pgfmathprintnumber[fixed,fixed zerofill,precision=0]{2.22}$	&	${11.6} \pm \pgfmathprintnumber[fixed,fixed zerofill,precision=0]{0.39}$
\\											&	R2d	&	&			
${81.1} \pm \pgfmathprintnumber[fixed,fixed zerofill,precision=0]{5.63}$	&	${90.6} \pm \pgfmathprintnumber[fixed,fixed zerofill,precision=0]{2.81}$	&	${14.7} \pm \pgfmathprintnumber[fixed,fixed zerofill,precision=0]{0.33}$
&	${82.5} \pm \pgfmathprintnumber[fixed,fixed zerofill,precision=0]{5.45}$	&	$\mathbf{91.2} \pm \pgfmathprintnumber[fixed,fixed zerofill,precision=0]{2.73}$	&	$\mathbf{14.6} \pm \pgfmathprintnumber[fixed,fixed zerofill,precision=0]{0.22}$
\\											&	R2e	&	&			
${57.0} \pm \pgfmathprintnumber[fixed,fixed zerofill,precision=0]{7.24}$	&	$\mathbf{78.5} \pm \pgfmathprintnumber[fixed,fixed zerofill,precision=0]{3.62}$	&	${18.5} \pm \pgfmathprintnumber[fixed,fixed zerofill,precision=0]{0.27}$
&	${56.7} \pm \pgfmathprintnumber[fixed,fixed zerofill,precision=0]{7.86}$	&	${78.3} \pm \pgfmathprintnumber[fixed,fixed zerofill,precision=0]{3.93}$	&	$\mathbf{17.5} \pm \pgfmathprintnumber[fixed,fixed zerofill,precision=0]{0.24}$
\\											&	R4	&	&			
${75.3} \pm \pgfmathprintnumber[fixed,fixed zerofill,precision=0]{3.22}$	&	$\mathbf{81.5} \pm \pgfmathprintnumber[fixed,fixed zerofill,precision=0]{2.41}$	&	${49.2} \pm \pgfmathprintnumber[fixed,fixed zerofill,precision=0]{0.55}$
&	${75.2} \pm \pgfmathprintnumber[fixed,fixed zerofill,precision=0]{3.90}$	&	${81.4} \pm \pgfmathprintnumber[fixed,fixed zerofill,precision=0]{2.93}$	&	$\mathbf{46.4} \pm \pgfmathprintnumber[fixed,fixed zerofill,precision=0]{0.40}$
\\
\hline
\end{tabular}
}
\end{table}
}

%% file: rules.tex
\newcommand{\textvar}[1]{\text{\uppercase{#1}}}
{
\begin{table}[tp]
\caption{Test coverage
and confidences for audio rules extracted on binary classification tasks E2, R2a, R2c, R2d, and R2e. For each task, all well-performing rules from  propositional and modal trees are shown.
Variable names (e.g., FLUX, FLATNESS, MFCC$_i$, etc.) are capitalized, and the corresponding Hertz frequencies are shown for variables representing Mel frequencies (e.g., MEL$_{2053\text{Hz}}$). Furthermore, Catch22 features are shown as
$\text{entropy}_{\text{pairs}}$, 
$\text{transition}_{\text{var}}$, 
$\text{stretch}_{\text{high}}$, and 
$\text{stretch}_{\text{decr}}$. 
}\label{rules}
\centering
\normalsize
\resizebox{0.98\linewidth}{!}{
\begin{tabular}{|l|r|l|r|r|}

\hline
\multicolumn{1}{|c|}{antecedent} &
\multicolumn{1}{|c|}{consequent} &
\multicolumn{1}{|c|}{cov.} &
\multicolumn{1}{|c|}{conf.} \\

\hline
\multicolumn{4}{|c|}{E2} \\ \hline

$(\text{median}(\textvar{mel}_{2316\text{Hz}}) < 0.09) \wedge (\text{mean}(\textvar{entropy}) < 1.80) \wedge (\text{std}(\textvar{mel}_{198\text{Hz}}) \ldots$
 & negative & 19 & 0.89 \\
$(\text{median}(\textvar{mel}_{2316\text{Hz}}) \geq 0.09) \wedge (\text{median}(\textvar{mfcc}_{5}) \geq 7.50) \wedge (\text{transition}_{\text{var}}(\ldots$
& positive & 23 & 0.82 \\
$(\text{std}(\textvar{mel}_{2581\text{Hz}}) \geq 477.76) \wedge (\text{std}(\textvar{mel}_{319\text{Hz}}) \geq 2.68\cdot 10^{-5}) \wedge (\text{std}(\textvar{mel}_{1457\text{Hz}}) \geq 1444.81) \wedge \ldots$
& negative & 21 & 0.66 \\
$(\text{std}(\textvar{mel}_{2581\text{Hz}}) < 477.76) \wedge (\text{stretch}_{\text{high}}(\textvar{mel}_{516\text{Hz}}) \geq -18.05) \wedge (\text{stretch}_{\text{high}}(\ldots$
& positive & 26 & 0.89 \\

\hline

$\exmodop{G}(\text{std}(\textvar{crest}) \leq 0.32) \wedge \univmodop{G}(\text{max}(\textvar{mel}_{227\text{Hz}}) < 0.77)$ & negative & 31 & 0.73 \\
$\univmodop{G}(\text{std}(\textvar{mel}_{4000\text{Hz}}) < 0.02 \wedge \text{std}(\textvar{mfcc}_{9}) < 0.39 \wedge \text{transition}_{\text{var}}(\textvar{mel}_{100\text{Hz}}) > 0.01)$ & positive & 22 & 0.81 \\

\hline
\multicolumn{4}{|c|}{R2a} \\ \hline


$(\text{std}(\textvar{mfcc}_{2}) < 2.46)$ & sick & 17 & 1.00 \\
$(\text{std}(\textvar{mfcc}_{2}) \geq 2.46)$ & healthy & 20 & 1.00 \\
$(\text{transition}_{\text{var}}(\textvar{mel}_{2756\text{Hz}}) \geq -0.92)$ & sick & 17 & 1.00 \\
$(\text{transition}_{\text{var}}(\textvar{mel}_{2756\text{Hz}}) < -0.92)$ & healthy & 18 & 1.00 \\


\hline


$\exmodop{G}(\text{max}(\textvar{decrease}) \leq -4.92)$ & healthy & 15 & 1.00 \\
$\exmodop{G}(\text{max}(\textvar{decrease}) \leq -4.55 \wedge \text{max}(\textvar{mel}_{353\text{Hz}}) \leq 0.01)$ & healthy & 21 & 1.00 \\


\hline

\multicolumn{4}{|c|}{R2b} \\ \hline


$(\text{median}(\textvar{mel}_{504\text{Hz}}) < 1.90) \wedge (\text{min}(\textvar{mel}_{3235\text{Hz}}) < 2.3\cdot 10^{-5})$ & healthy & 10 & 1.00 \\
$(\text{median}(\textvar{mel}_{504\text{Hz}}) \geq 1.90) \wedge (\text{entropy}_{\text{pairs}}(\textvar{mel}_{2405\text{Hz}}) \geq -1.16) \wedge (\text{stretch}_{\text{decr}}(\textvar{mel}_{1750\text{Hz}}) \geq 0.18)$ & sick & 14 & 0.79 \\
$(\text{min}(\textvar{mel}_{393\text{Hz}}) \geq 1.46) \wedge (\text{stretch}_{\text{high}}(\textvar{mel}_{2756\text{Hz}}) < 924.73)$ & sick & 12 & 0.83 \\

\hline



$\exmodop{G}(\text{max}(\textvar{mfcc}_{5}) \geq 1.16 \wedge \exmodop{\overline{DBE}}(\text{min}(\textvar{mel}_{353\text{Hz}}) \geq 5.1\cdot 10^{-5} \wedge \exmodop{AO}(\text{min}(\textvar{mfcc}_{6}) \geq 0.51))) \ldots$ & sick & 18 & 0.72 \\
$\univmodop{G}(\text{max}(\textvar{mfcc}_{5}) < 1.26 \wedge \text{max}(\textvar{mel}_{353\text{Hz}}) < 0.01)$ & healthy & 11 & 1.00 \\
$\exmodop{G}(\text{max}(\textvar{mel}_{2053\text{Hz}}) \leq 0.01 \wedge \exmodop{\overline{L}}(\text{std}(\textvar{entropy}) \leq 0.01)) \wedge \univmodop{G}(\text{max}(\textvar{mfcc}_{3}) < 0.98)$ & healthy & 13 & 0.85 \\
$\exmodop{G}(\text{max}(\textvar{mfcc}_{3}) \geq 0.98 \wedge \text{mean}(\textvar{skewness}) \leq 6.25)$ & sick & 14 & 1.00 \\

\hline

\multicolumn{4}{|c|}{R2c} \\ \hline


$(\text{mean}(\textvar{mel}_{1750\text{Hz}}) < 0.03) \wedge (\text{min}(\textvar{mel}_{2595\text{Hz}}) < 0.36)$ & sick & 14 & 0.86 \\
$(\text{mean}(\textvar{mel}_{1750\text{Hz}}) \geq 0.03) \wedge (\text{stretch}_{\text{decr}}(\textvar{mel}_{919\text{Hz}}) < -0.36) \wedge (\text{mean}(\textvar{mel}_{2796\text{Hz}}) < 0.18)$ & healthy & 19 & 0.95 \\
$(\text{mean}(\textvar{mel}_{681\text{Hz}}) \geq 1.8\cdot 10^{-6}) \wedge (\text{min}(\textvar{mel}_{609\text{Hz}}) < 68.08) \wedge (\text{mean}(\textvar{mel}_{541\text{Hz}}) \geq 0.61)$ & sick & 16 & 0.88 \\
$(\text{mean}(\textvar{mel}_{681\text{Hz}}) < 1.8\cdot 10^{-6}) \wedge (\text{min}(\textvar{mel}_{2179\text{Hz}}) \geq 3.2\cdot 10^{-9}) \wedge (\text{stretch}_{\text{decr}}(\textvar{mel}_{837\text{Hz}}) < -0.27)$ & healthy & 26 & 0.81 \\
$(\text{stretch}_{\text{high}}(\textvar{mel}_{733\text{Hz}}) \geq 1.35)$ & sick & 11 & 1.00 \\
$(\text{stretch}_{\text{high}}(\textvar{mel}_{733\text{Hz}}) < 1.35) \wedge (\text{entropy}_{\text{pairs}}(\textvar{mel}_{1541\text{Hz}}) \geq 2.1\cdot 10^{-6}) \wedge (\text{median}(\textvar{mel}_{393\text{Hz}}) \ldots$ & healthy & 24 & 0.79 \\


\hline


$\exmodop{G}(\text{max}(\textvar{mfcc}_{6}) \leq 0.23 \wedge \exmodop{AO}(\text{min}(\textvar{mfcc}_{7}) \geq 0.01)) \wedge \univmodop{G}(\text{max}(\textvar{mel}_{734\text{Hz}}) > 9.3\cdot 10^{-6} \ldots$ & sick & 21 & 1.00 \\
$\univmodop{G}(\text{max}(\textvar{mel}_{809\text{Hz}}) > 8.9\cdot 10^{-6} \wedge \text{max}(\textvar{mfcc}_{2}) > 2.17 \wedge \text{std}(\textvar{mel}_{1065\text{Hz}}) < 0.00)$ & sick & 13 & 0.85 \\
$\exmodop{G}(\text{max}(\textvar{mel}_{809\text{Hz}}) \leq 8.9\cdot 10^{-6} \wedge \text{max}(\textvar{mfcc}_{3}) \leq 0.63) \wedge \univmodop{G}(\text{max}(\textvar{mel}_{809\text{Hz}}) \leq 8.9\cdot 10^{-6} \ldots$ & healthy & 27 & 1.00 \\
$\exmodop{G}(\text{std}(\textvar{mel}_{3355\text{Hz}}) \leq 0.00) \wedge \univmodop{G}(\text{max}(\textvar{mfcc}_{5}) < 0.93)$ & sick & 12 & 0.92 \\
$\exmodop{G}(\text{max}(\textvar{mfcc}_{5}) \geq 0.93 \wedge \text{median}(\textvar{flatness}) \geq 5\cdot 10^{-6} \wedge \exmodop{L}\text{max}(\textvar{mfcc}_{4}) \geq 0.88)$ & healthy & 20 & 1.00 \\


\hline

\multicolumn{4}{|c|}{R2d} \\ \hline


$(\text{stretch}_{\text{decr}}(\textvar{mel}_{919\text{Hz}}) \geq -0.50) \wedge (\text{stretch}_{\text{high}}(\textvar{mel}_{919\text{Hz}}) \geq 1.73) \wedge (\text{entropy}_{\text{pairs}}(\textvar{mfcc}_{8}) < 8.50)$ & sick & 16 & 1.00 \\
$(\text{stretch}_{\text{decr}}(\textvar{mel}_{919\text{Hz}}) < -0.50) \wedge (\text{median}(\textvar{mel}_{504\text{Hz}}) \geq 1.26) \wedge (\text{stretch}_{\text{decr}}(\textvar{mel}_{504\text{Hz}}) \geq 0.84)$ & healthy & 25 & 0.92 \\
$(\text{entropy}_{\text{pairs}}(\textvar{mel}_{2179\text{Hz}}) < 0.44) \wedge (\text{entropy}_{\text{pairs}}(\textvar{mel}_{1438\text{Hz}}) \geq -0.08) \wedge (\text{mean}(\textvar{mel}_{3648\text{Hz}}) \geq 0.44)$ & healthy & 20 & 1.00 \\
$(\text{entropy}_{\text{pairs}}(\textvar{mel}_{2179\text{Hz}}) \geq 0.44) \wedge (\text{std}(\textvar{mel}_{1438\text{Hz}}) < 0.40) \wedge (\text{stretch}_{\text{high}}(\textvar{mel}_{837\text{Hz}}) \geq 1.68)$ & sick & 23 & 1.00 \\


\hline


$\univmodop{G}(\text{max}(\textvar{mfcc}_{3}) > 0.17 \wedge \text{max}(\textvar{mfcc}_{6}) < 0.80 \wedge \text{entropy}_{\text{pairs}}(\textvar{mel}_{3083\text{Hz}}) < 2.22)$ & sick & 17 & 0.82 \\
$\exmodop{G}(\text{max}(\textvar{mfcc}_{3}) \leq 0.17 \wedge \exmodop{\overline{AO}}(\text{mean}(\textvar{mfcc}_{5}) \geq 0.22))$ & healthy & 23 & 1.00 \\
$\exmodop{G}(\text{max}(\textvar{mfcc}_{2}) \leq 2.24) \wedge \univmodop{G}(\text{max}(\textvar{mfcc}_{5}) < 0.81)$ & healthy & 13 & 1.00 \\
$\exmodop{G}(\text{max}(\textvar{mfcc}_{5}) \geq 0.81 \wedge \text{mean}(\textvar{mfcc}_{3}) \leq 0.70)$ & healthy & 18 & 1.00 \\
$\univmodop{G}(\text{max}(\textvar{mfcc}_{5}) < 0.81 \wedge \text{max}(\textvar{mfcc}_{2}) > 2.24 \wedge \text{max}(\textvar{mfcc}_{5}) < 0.69)$ & sick & 20 & 0.95 \\


\hline
\multicolumn{4}{|c|}{R2e} \\ \hline


$(\text{min}(\textvar{mel}_{2595\text{Hz}}) \geq 0.50) \wedge (\text{max}(\textvar{mel}_{2595\text{Hz}}) \geq 0.38) \wedge (\text{min}(\textvar{mel}_{504\text{Hz}}) < 2.65) \ldots$ & sick & 10 & 1.00 \\
$(\text{min}(\textvar{mel}_{2595\text{Hz}}) < 0.50) \wedge (\text{transition}_{\text{var}}(\textvar{flux}) < 1.74) \wedge (\text{stretch}_{\text{decr}}(\textvar{flatness}) < 2.20) \ldots$ & healthy & 13 & 0.77 \\
$(\text{max}(\textvar{mel}_{2368\text{Hz}}) < 0.86) \wedge (\text{stretch}_{\text{high}}(\textvar{mel}_{681\text{Hz}}) < 1.5\cdot 10^{-6}) \wedge (\text{std}(\textvar{mel}_{2006\text{Hz}}) \geq 7.8\cdot 10^{-7})$ & healthy & 10 & 0.90 \\
$(\text{max}(\textvar{mel}_{2368\text{Hz}}) \geq 0.86) \wedge (\text{std}(\textvar{mel}_{2575\text{Hz}}) \geq 86.70) \wedge (\text{transition}_{\text{var}}(\textvar{mel}_{1438\text{Hz}}) \geq -0.19)$ & sick & 13 & 1.00 \\
$(\text{median}(\textvar{mel}_{393\text{Hz}}) \geq 1.02) \wedge (\text{max}(\textvar{mel}_{820\text{Hz}}) \geq -0.93) \wedge (\text{median}(\textvar{mfcc}_{2}) < 10.26) \ldots$ & sick & 16 & 0.86 \\


\hline


$\univmodop{G}(\text{max}(\textvar{mfcc}_{2}) > 2.98 \wedge \text{max}(\textvar{mel}_{1161\text{Hz}}) > 1.2\cdot 10^{-7} \wedge \text{max}(\textvar{mfcc}_{10}) > -0.74 \ldots$ & sick & 8 & 1.00 \\
$\exmodop{G}(\text{max}(\textvar{mfcc}_{2}) \leq 2.98) \wedge \exmodop{G}(\text{max}(\textvar{mel}_{890\text{Hz}}) \leq 5.8\cdot 10^{-6}) \wedge [G](\text{max}(\textvar{mfcc}_{2}) \leq 2.98 \ldots$ & healthy & 10 & 0.90 \\
$\exmodop{G}(\text{std}(\textvar{mel}_{1373\text{Hz}}) \geq 1.8\cdot 10^{-6} \wedge \exmodop{\overline{AO}}(\text{max}(\textvar{mfcc}_{6}) \leq -0.76))$ & healthy & 11 & 0.82 \\
$\exmodop{G}(\text{std}(\textvar{mel}_{1373\text{Hz}}) \geq 1.8\cdot 10^{-6} \wedge \exmodop{AO}(\text{max}(\textvar{mfcc}_{5}) \geq 1.26)) \wedge \univmodop{G}(\text{std}(\textvar{mel}_{1373\text{Hz}}) \ldots$ & sick & 14 & 0.86 \\


\hline

\end{tabular}
}
\end{table}
}